\documentclass[pra]{revtex4}
\usepackage{amsmath}
\usepackage{epsfig}
\usepackage{array}
\usepackage{natbib}

\begin{document}

\title{Enhancement of the electron electric dipole moment in
gadolinium garnets}

\author{T. N. Mukhamedjanov, V. A. Dzuba, O. P. Sushkov}

\affiliation{School of Physics, University of New South Wales,\\
 Sydney 2052, Australia}

\begin{abstract}
Effects caused by the electron electric dipole moment (EDM) in gadolinium
garnets are considered. Experimental studies of these effects could improve
current upper limit on the electron EDM by several orders of magnitude.
We suggest a consistent theoretical model and perform calculations
of observable effects in gadolinium gallium garnet and gadolinium
iron garnet. Our calculation accounts for both direct and exchange diagrams.
\end{abstract}

\maketitle

\section{introduction}

Violation of the combined symmetry of charge conjugation (C) and parity (P)
has been discovered in the decay of the $K^0$ meson about 40 years ago
\cite{CCK}. The exact origin of this symmetry violation remains an enigma,
although the so-called standard model of electroweak interactions can
describe these processes phenomenologically. It has also been proposed by
Sakharov \cite{Sak} that the matter--antimatter asymmetry observed in our
Universe could have arisen from a CP-violating interaction active at an
early stage of the Big Bang. The CP-violation implies a time-reversal (T)
asymmetry, because there are strong reasons to  believe that the combined
CPT-symmetry should not be violated \cite{CPT}.
Violation of the time reversal symmetry has been observed recently in decays
of K-mesons \cite{KK}.
T-violation together with well known parity (P) violation
(T,P-violation) provides a nonzero electric dipole moment (EDM) of a
system in a stationary quantum state. This is why searches for EDM of
elementary particles, atoms and molecules are very important for
studies of  violations of fundamental symmetries \cite{KL}.
In the present work we concentrate on the electron EDM.
The present best limitation on the electron EDM comes from the
experiment with an atomic Thallium beam \cite{Com},
\begin{equation}
\label{Tl}
d_e < 1.6\times 10^{-27}e\cdot cm.
\end{equation}
There have been recent suggestions \cite{Lam,Hun} to improve
the sensitivity to the electron EDM substantially by working with
solids. In particular gadolinium gallium garnet,
Gd$_3$Ga$_5$O$_{12}$ (GGG), and gadolinium iron garnet
Gd$_3$Fe$_5$O$_{12}$ (GdIG) have been suggested.
The idea for searches of the electron EDM in solid state
experiments was first suggested by Shapiro in 1968 \cite{Sh}.
There are two ways to perform such an experiment.
The first one is to apply an external magnetic field to the solid.
Polarization of electrons by the magnetic field causes alignment
of electron EDMs, and hence induces a voltage across the sample that could
be detected. Another possibility is to apply a strong external electric
field. This would align the EDMs of bound electrons and hence
lead to a simultaneous alignment of the electron spins; the
magnetic field arising from this alignment could be detected
experimentally.
An experiment of this kind has been performed with
nickel-zinc ferrite \cite{VK}. Due to experimental
limitations the result was not very impressive.
However, according to estimates presented in Refs. \cite{Lam,Hun},
application of novel experimental techniques and
measurements with GGG and GdIG makes this direction  highly promising.

The first calculation of the T,P-odd effects in GGG and GdIG
has been performed in Ref.~\cite{Simon} (see also Refs.
\cite{Stefan, Volodya}).
The mechanism considered in that work was similar to the mechanism of
T,P-violation in atoms due to the T,P-odd nuclear forces \cite{SFK}.
In essence the Gd$^{3+}$ ion has been treated as a large nucleus that has
some effective Schiff moment. External electrons that belong to O$^{2-}$
ions penetrate inside the Gd$^{3+}$ ion and interact with the ``Schiff
moment'' of the ion.
This results in an effective interaction between the lattice deformation and
the electron EDM. It has been also pointed out in Ref.~\cite{Simon}
that there is a contribution that can not be reduced to the ``Schiff
moment mechanism'', this contribution is due to the exchange between
external and internal electrons. However the exchange diagrams
estimated in Ref.~\cite{Simon} gave a small contribution.
In the present work we have found a new class of exchange diagrams
that are very important. Calculation of the effect with account of
exchange diagrams is not a simple problem because in this case the
logic that leads to the Schiff moment \cite{SFK,Simon} is not valid. 
For the same reason the simplistic model of the electronic structure
of Gadolonium garnet used in \cite{Simon} is not sufficient for the
present calculation.
Therefore to perform the present calculation we have developed
a more accurate model that treats gadolinium and oxygen electrons simultaneously.
We believe that the present calculation is more accurate than that
performed in Ref.~\cite{Simon}. Nevertheless, surprisingly, the results
are very close.

\section{theoretical model for electronic structure}

\subsection{Single-electron effective potential}

Compounds of our interest, gadolinium iron garnet and gadolinium gallium
garnet, are ionic crystals,
consisting of Gd$^{3+}$, Fe$^{3+}$ (or Ga$^{3+}$) and O$^{2-}$ ions.
Uncompensated electron spins are localized at Gd$^{3+}$ ion, which has a
$4f^7$ configuration, and Fe$^{3+}$ ion, whose electronic configuration is
$4d^5$. Both  paramagnetic ions contribute to the T,P-odd effect.
However, contribution of the ion to the T,P-odd effect scales as $Z^3$,
where $Z$ is the nuclear charge (see, e.g.~\cite{KL}). Therefore, we
neglect the contribution of Fe$^{3+}$ and consider only Gd$^{3+}$ ions.

To describe an isolated Gd$^{3+}$ ion we use the effective potential
in the following parametric form
\begin{eqnarray}
\label{pot}
V_{Gd}(\boldsymbol{r})&=&\frac{1}{r}\frac{(Z_i-Z)
(e^{-\frac{\mu}{d}}+1)}
{(1+\eta r)^2(e^{\frac{r-\mu}{d}}+1)}-\frac{Z_i}{r}.
\end{eqnarray}
Here we use atomic units, $Z$ is charge of the nucleus, $Z_i$ is charge of
the core of the ion, and $\mu$, $d$ and $\eta$ are some parameters that 
describe the core. Solution of the Dirac equation with the potential (\ref{pot})
gives wave functions and energies of electron states.
We use the following values of the parameters 
\begin{eqnarray}
\mu = 1.00, \ \ \ d = 1.00 \ \ \ \eta = 2.35 \ .
\end{eqnarray}
These values  provide a fit of experimental energy levels for Gd$^{3+}$ 
($Z_i=4$) and Gd$^{2+}$ ($Z_i=3$).
Calculated and  experimental \cite{Atdat} energy levels
averaged over fine structure are shown in Table \ref{tab1}.

\begin{table}[h!]
\centering
\begin{tabular}{l c c c}
\hline \hline
 Ion & electron state  & Calculaton & Experiment \\ 
\hline \hline
Gd$^{3+}$& $4f$ & -363 & -355 \\
\cline{2-4} 
  & $5d$ & -281 &   \\ 
\hline \hline
 Gd$^{2+}$ & $5d$ & -156 & -157 \\ 
\cline{2-4}
  & $6s$ & -120 & -121 \\ 
\cline{2-4}
  & $6p$ & -158 & -156 \\ 
\hline \hline
\end{tabular}
\caption{
Calculated and experimental energy levels with respect to the ionization limit.
The levels are averaged over fine structure. Units $10^3 \ cm^{-1}$.
}
\label{tab1}
\end{table}

\subsection{Account of Gd$\boldsymbol{^{3+}}$ environment}

\begin{figure}[!hbt]
\centering
\epsfig{figure=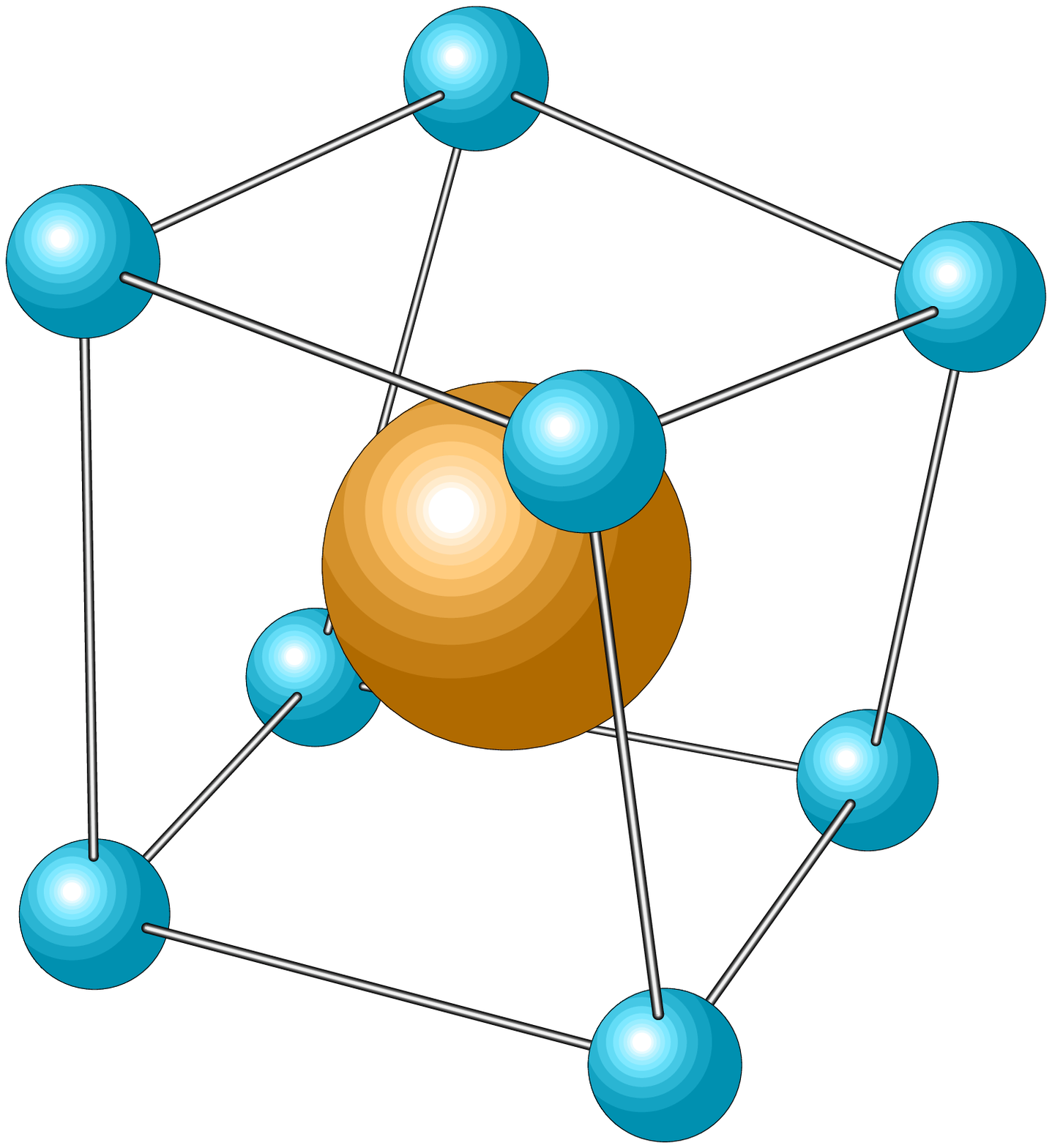,height=4cm}\qquad\qquad
\epsfig{figure=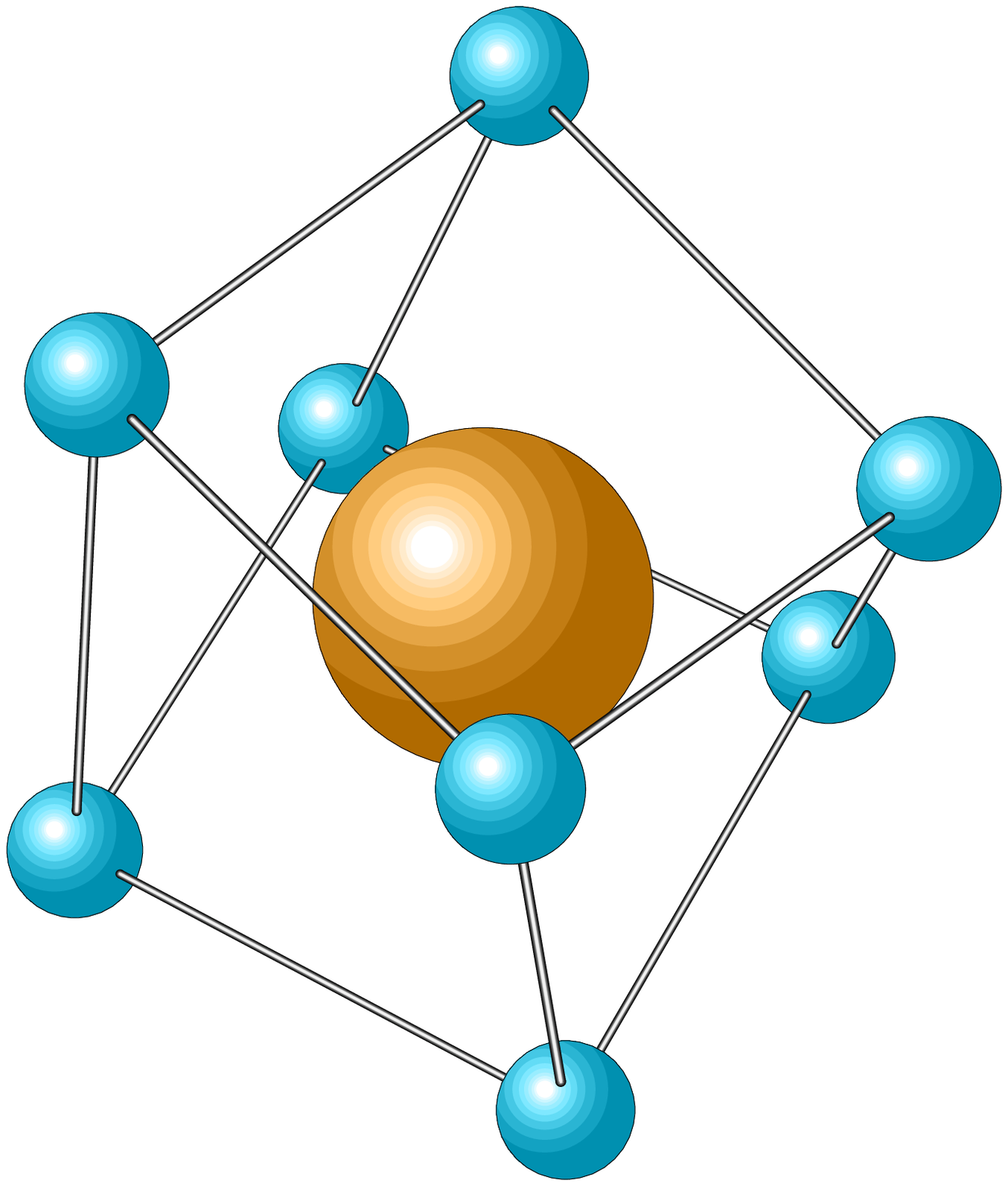,height=4.2cm}
\caption{Dodecahedron configuration of O$^{2-}$ ions around Gd$^{3+}$ ion in 
garnet structure.}
\label{Fig1}
\end{figure}

\begin{figure}[h]
\centering
\includegraphics[height=120pt,keepaspectratio=true]{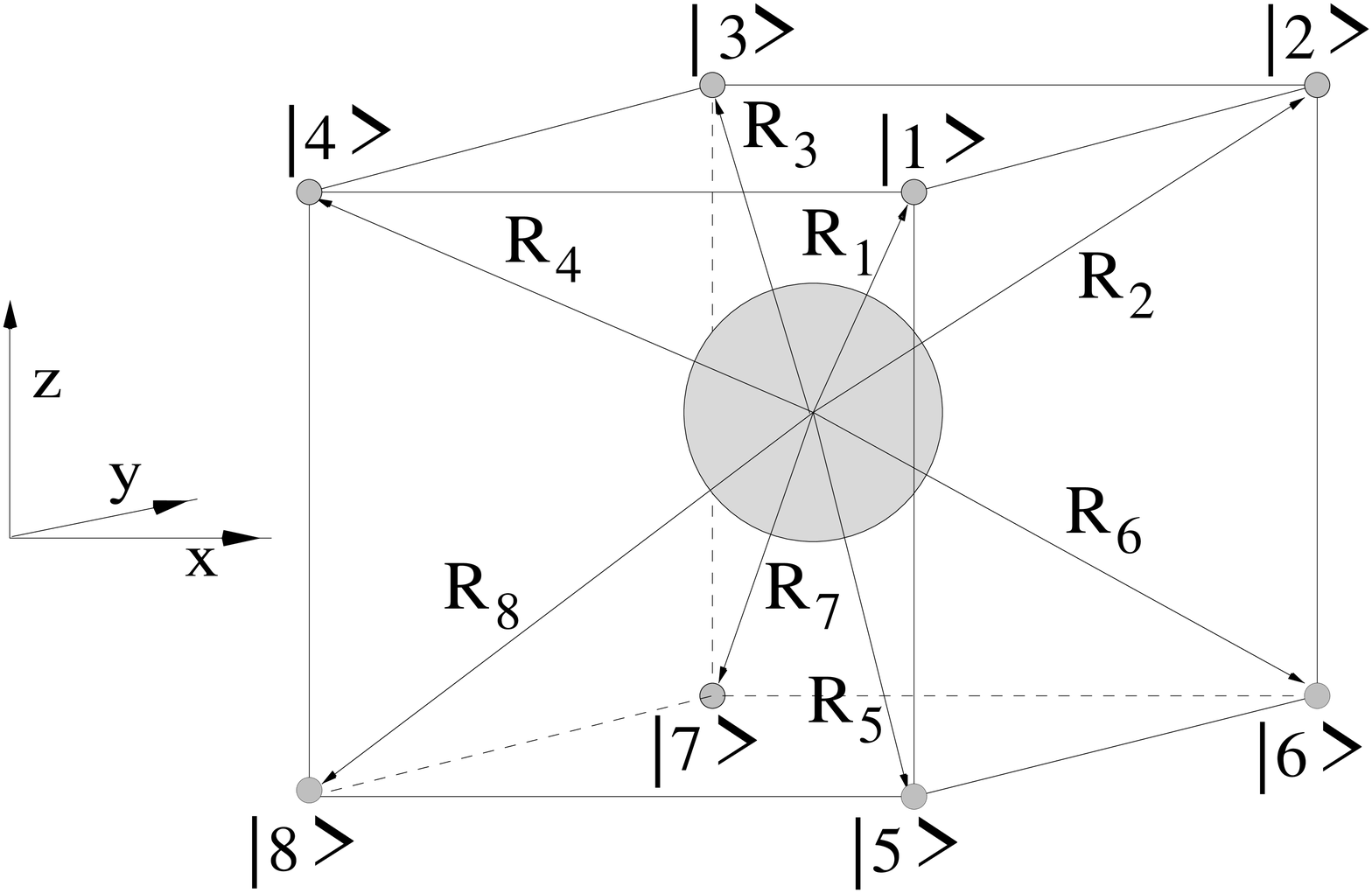}
\caption{\it 
Gd$^{3+}$ ion with surrounding eight oxygen ions.}
\label{FigC}
\end{figure}

In the garnet structure each gadolinium ion is surrounded by eight oxygen ions 
O$^{2-}$ in a dodecahedron configuration (resembles distorted cube),
see Ref.~\cite{GGG}. The configuration at two different angles of view is
shown in  Fig.~\ref{Fig1}.
The Gd$-$O distance is $r_o=4.53a_B$, where $a_B$ is the Bohr radius. We 
need to know wave functions of  oxygen electrons inside the Gd$^{3+}$ ion. 
Electronic configuration of O$^{2-}$ is
$1s^22s^22p^6$. Consider 2p orbitals of the ion.
It can be shown that $2p_\pi$ orbitals do not contribute to the effect and we
need to consider only $2p_\sigma$ orbitals, pointing towards Gd$^{3+}$. Every
$2p_{\sigma}$ orbital is double occupied, so we have 16 electrons in the 
vicinity of Gd$^{3+}$.
In first approximation one can neglect distortion of the oxygen cube
and use a linear combination of oxygen $2p_{\sigma}$ orbitals with definite symmetry with respect to the cubic group.
We are particularly interested in the $|S\rangle$ wavefunction, which is 
symmetric with respect to reflection of cube's main axes, and $|P_x\rangle$, 
$|P_y\rangle$ and $|P_z\rangle$ wavefunctions, which change sign with 
reflection of $x$, $y$ and $z$ axes, correspondingly. They are of the form:
\begin{eqnarray}
\label{SP}
|S \rangle \;\: &=&
\frac{ |1\rangle + |2\rangle + |3\rangle + |4\rangle
+ |5\rangle + |6\rangle + |7\rangle + |8\rangle }{\sqrt{8}}\ ,
\nonumber\\
|P_x \rangle &=&
\frac{ |1\rangle + |2\rangle - |3\rangle - |4\rangle
+ |5\rangle + |6\rangle - |7\rangle - |8\rangle }{\sqrt{8}}\ ,
\nonumber\\
|P_y \rangle &=&
\frac{  |1\rangle - |2\rangle - |3\rangle + |4\rangle
+ |5\rangle - |6\rangle - |7\rangle + |8\rangle}{\sqrt{8}}\ ,
\nonumber\\
|P_z \rangle &=&
\frac{ |1\rangle + |2\rangle + |3\rangle + |4\rangle
- |5\rangle - |6\rangle - |7\rangle - |8\rangle}{\sqrt{8}}\ .
\end{eqnarray}
Here $|n\rangle$ denotes $2p_\sigma$ orbital of $n$-th oxygen ion as they 
are enumerated in Fig.~\ref{FigC}. There are also three D-wave states and 
one F-wave state combined from oxygen $2p_\sigma$ orbitals, see Ref.~\cite{Simon},
but these states do not contribute to the effect and we neglect them.
In the work \cite{Simon} the wave functions (\ref{SP}) have been matched
to the gadolinium $6s$ and $6p$ wave functions at the matching sphere of
radius $R \approx 2.5a_B$ around Gd. This is a simple and reliable
way to describe penetration of oxygen electrons inside Gd$^{3+}$.
However, unfortunately this is not a consistent quantum mechanical
description because the space is divided in two parts --- inside and outside
of the matching sphere. Our analysis has shown that calculation of exchange
diagrams in this simple model is ambiguous. The point is that the effect
arises in the third order of perturbation theory (see below). In such a high order
one must use a consistent quantum mechanical description.
The best solution of the problem would be a 3-dimensional Hartree-Fock
or 3-dimensional pseudo-potential calculation for the cluster consisting of
Gd and eight oxygen ions (GdO$^8$-cluster). However this is a very complex calculation.
In the present work we suggest an intermediate solution, essentially a
jelly model. We assume that eight oxygen ions provide a spherically symmetric
attractive potential for electrons
\begin{equation}
V_{O}(\boldsymbol{r})=-A_o e^{-\left(\frac{r-r_o}{D}\right)^2}
\label{pot1}
\end{equation}
where $r_o=4.53$ is Gd-O distance, and $A_o$ and  $D$ are some parameters.
So effectively we replace the dodecahedron in Fig.~\ref{Fig1} or the cube in
Fig.~\ref{FigC} by a sphere. Thus the electrons are moving in the combined
spherically symmetric potential
\begin{equation}
\label{pot2}
V(\boldsymbol{r})=V_{Gd}(\boldsymbol{r})+V_O(\boldsymbol{r}),
\end{equation}
where $V_{Gd}$ is the potential of Gd$^{3+}$ core (\ref{pot}), and $V_O$ is the
combined potential of oxygens  (\ref{pot1}). 
Solution of the Dirac equation with potential (\ref{pot2}) gives single
particle orbitals. In this picture  the electronic configuration of the
GdO$^8$-cluster is $1s^2\! ...5s^25p^64f^76s^26p^6$. Out of this
$1s^2\! ... 5s^25p^64f^7$ are the Gd$^{3+}$ electrons and $6s^26p^6$ are in essence
the oxygen electrons. So we have eight oxygen electrons, this is exactly what one
needs to describe S and P $2p_{\sigma}$ states (\ref{SP}).
In this model one cannot describe D and  F states combined from
$2p_{\sigma}$ oxygen orbitals, but fortunately there is no need in these
states because they do not contribute to the effect.

\begin{figure}[h]
\vspace{10pt}
\centering
\includegraphics[height=120pt,keepaspectratio=true]{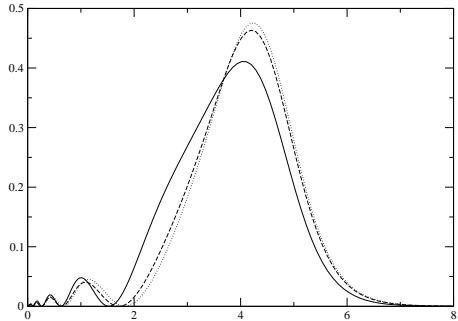}
\vspace{10pt}
\caption{\it 
Radial electron density $|f(r)|^2$, $\psi(r)=f(r)/r\cdot Y_{lm}(\theta,\phi)$,
versus distance from the Gd nucleus. The solid line corresponds to the $6s$-state,
the dashed line corresponds to the $6p_{1/2}$-state, and 
the dotted line corresponds to the $6p_{3/2}$-state.}
\label{Fig6sp}
\end{figure}

How to determine the constants $A_o$ and $D$ in the effective oxygen
``jelly'' potential (\ref{pot1})? The size of the potential $D$ is not very 
important, it is clear that $D\sim 1$ (we remind that we use atomic units),
and we set $D=1$. The depth of the potential $A_o$ is crucially important.
Wave functions of oxygen ion O$^{2-}$ have been calculated in
Ref.~\cite{FS},
so $\psi_{2p_{\sigma}}$ is known. 
We chose the depth $A_o$ from the requirement that the wave functions
$\psi_{6s}$ and $\psi_{6p}$ satisfy the following conditions
\begin{eqnarray}
\label{dual}
&&|\psi_{6s}(R)|=|\psi_{2p_{\sigma}}(r_o-R,\cos\theta=1)|\ ,\nonumber\\
&&|\psi_{6p}(R,\cos\theta=1/\sqrt{3})|=
|\psi_{2p_{\sigma}}(r_o-R,\cos\theta=1)|\ ,
\end{eqnarray}
where $R\approx 2.5$. This is just an alternative formulation of the
idea of dual description at $r \approx R$, see Refs. \cite{FS,Simon}.
There are two conditions (\ref{dual}) and only one parameter $A_o$,
so strictly speaking one cannot satisfy both conditions exactly.
However at $A_o=0.9$ each of the conditions (\ref{dual}) is satisfied
with accuracy $\sim 15\%$, and this is the value of $A_o$ which we use in our 
calculations. Thus, parameters of the ``oxygen'' potential (\ref{pot1}) are
\begin{eqnarray}
r_o=4.53, \ \ \ A_o=0.9, \ \ \ D=1.
\end{eqnarray}
Let $f(r)$ be the upper component of the Dirac spinor, 
$\psi(r)=f(r)/r\cdot Y_{lm}(\theta,\phi)$. The electron density $|f(r)|^2$ 
for $6s$, $6p_{1/2}$, and $6p_{3/2}$ states calculated in the potential
(\ref{pot2}) is shown in Fig.~\ref{Fig6sp}.
The density is peaked at $r=r_o=4.53$, this is the effective description of
oxygen ions. Energies of 4f, 5d, 6s, and 6p states in the potential (\ref{pot2})
are listed in Table \ref{tab2}. The energies are given with respect to the ionization limit.
\begin{table}[h!]
\centering
\begin{tabular}{cccccccc}
\hline
\hline
State & $4f_{5/2}$ & $4f_{7/2}$ & $5d_{3/2}$ & $5d_{5/2}$ & $6s$ & $6p_{1/2}$ & $6p_{3/2}$ \\ 
\hline Energy & -366 & -360 & -299 & -295 & -311 & -289 & -286 \\ 
\hline
\hline 
\end{tabular}\\
\caption{
Energy levels in the potential (\ref{pot2}) with respect to the ionization limit.
Units $10^3 \ cm^{-1}$.
}
\label{tab2}
\end{table}
It is worth to note that according to this calculation the 5d-4f splitting is
$\Delta E_{fd} \approx 66 \times 10^3cm^{-1}$. This is different from the splitting
for an isolated Gd$^{3+}$ ion, $\Delta E_{fd} \approx 100 \times 10^3cm^{-1}$, see
Ref.~\cite{Volodya}. However, the  value $\Delta E_{fd} \approx 66 \times 10^3cm^{-1}$
agrees well with that extrapolated for the Gd$^{3+}$ ion in the garnet environment, 
see Ref.~\cite{PDorenb}.

\section{T,P-odd energy correction related to the lattice deformation}

Similar to the approach \cite{Simon} in this section we consider an external 
deformation of the lattice consisting in a shift of the Gd$^{3+}$ ion with
respect to surrounding oxygen ions. Later we will relate the deformation
to observable effects.

We have already mentioned the T,P-odd effect arises in the third order
of perturbation theory. So, there are three perturbation theory operators 
we have to consider.
First of all this is  the operator of the T,P-odd interaction of the electron EDM 
$d_e$ with atomic electric field $\boldsymbol{E}$, see e.g.~Ref.~\cite{KL},
\begin{equation}
V_d=-d_e \gamma_o \boldsymbol{\Sigma\cdot E} \ .
\label{TPodd}
\end{equation}
Here $\gamma_o$ and $\boldsymbol{\Sigma}=\gamma_o\gamma_5\boldsymbol{\gamma}$ 
are Dirac $\gamma$-matrices. Because of the Schiff's theorem \cite{Sch} it is 
crucial to account for complex many-body screening effects, when working 
with the Hamiltonian (\ref{TPodd}). Technically this means that the many-body 
perturbation theory with operator (\ref{TPodd}) is very poorly convergent.
The standard way \cite{KL} to avoid this complication is to split the Hamiltonian 
(\ref{TPodd}) into two terms: $V_d=-d_e 
\gamma_o \boldsymbol{\Sigma\cdot E} = -d_e \boldsymbol{\Sigma\cdot E} -d_e 
(\gamma_o - 1) \boldsymbol{\Sigma\cdot E}$. Then due to the Schiff's theorem 
 contribution of the first term to an observable effect is 
identically zero, so one can reduce the interaction
\begin{equation}
V_d\rightarrow V^r_d=-d_e (\gamma_o - 1) \boldsymbol{\Sigma\cdot E}.
\label{pert1}
\end{equation}
The perturbation theory expansion with this operator is reasonably convergent.

\begin{figure}[!hbt]
\centering
\epsfig{figure=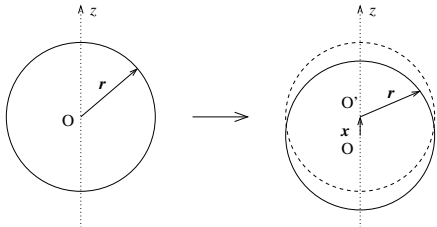,height=3cm}
\caption{Schematic picture, illustrating the shift of $V_O(\boldsymbol{r})$
due to the lattice deformation.}
\label{Figshift}
\end{figure}

The second perturbation operator is related to the shift of the 
Gd$^{3+}$ ion with respect to the surrounding oxygen ions. Let us denote value
of the shift by $x$. In our model, this corresponds to the shift of spherically 
symmetric $V_O(r)$ (\ref{pot1})  along one of the axes, say $z$-axis, see
Fig.~\ref{Figshift}

\noindent
Therefore
\begin{equation}
V_O(\boldsymbol{r}) \rightarrow V^{'} _O(\boldsymbol{r})=V_O(\boldsymbol{r}+ 
\boldsymbol{x})= V_O(\boldsymbol{r})+\frac{(\boldsymbol{x\cdot r)}}{r} 
\frac{\partial V_O}{\partial r}=V_O(\boldsymbol{r})+x \cos\theta \frac{\partial 
V_O}{\partial r} \ .
\end{equation}
We keep the term linear in $x$, $\theta$ is the azimuthal angle.
Thus the perturbation operator related to the lattice deformation reads
\begin{equation}
V_x(r)=x \cos\theta \frac{\partial V_O}{\partial r}=-2x \cos \theta 
\frac{(r-r_o)}{D^2}V_O(r)
\label{pert2}
\end{equation}

The third perturbation operator is residual electron-electron Coulomb interaction, which is not 
included in the effective potential
\begin{equation}
\label{pert3}
V_C(\boldsymbol{r}_i,\boldsymbol{r}_j)=\frac{1}{|\boldsymbol{r}_i -
\boldsymbol{r}_j|}=
\sum_{lm}\frac{4\pi}{2l+1}\frac{r_<^l}{r_>^{l+1}}
Y^*_{lm}(\boldsymbol{r}_i) Y_{lm}(\boldsymbol{r}_j)
\end{equation}
Here $\boldsymbol{r}_i$ and $\boldsymbol{r}_j$ are radius-vectors of interacting 
electrons.

\begin{figure}[h!]

\begin{tabular}{c c c c c c}

\multicolumn{1}{m{1cm}}{$\quad\quad\! 4 \times$} &
\multicolumn{1}{m{3.6cm}}{\epsfig{figure=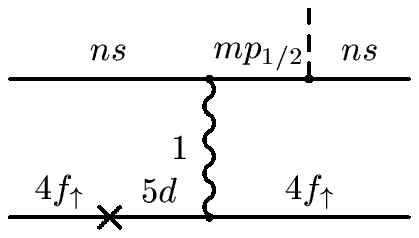,height=2cm}}&
\multicolumn{1}{m{1cm}}{$+\quad 4 \times$} &
\multicolumn{1}{m{3.6cm}}{\epsfig{figure=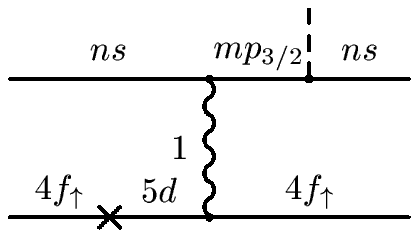,height=2cm}}&
\multicolumn{1}{m{1cm}}{$+\quad 4 \times$} &
\multicolumn{1}{m{3.6cm}}{\epsfig{figure=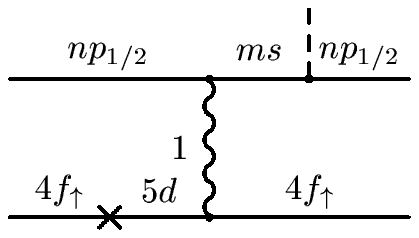,height=2cm}}\\
&1)& &2)& &3)\\

\multicolumn{1}{m{1cm}}{$+\quad 4 \times$} &
\multicolumn{1}{m{3.6cm}}{\epsfig{figure=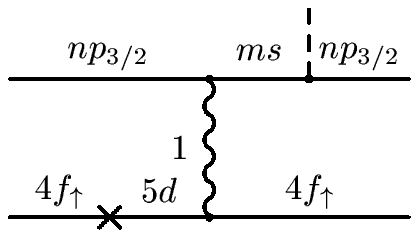,height=2cm}}&
\multicolumn{1}{m{1cm}}{$-\quad 2 \times$} &
\multicolumn{1}{m{3.6cm}}{\epsfig{figure=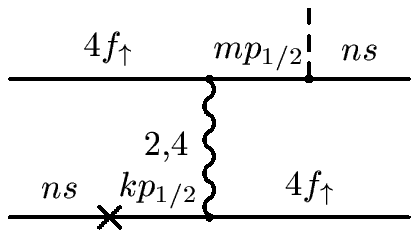,height=2cm}}&
\multicolumn{1}{m{1cm}}{$-\quad 2 \times$} &
\multicolumn{1}{m{3.6cm}}{\epsfig{figure=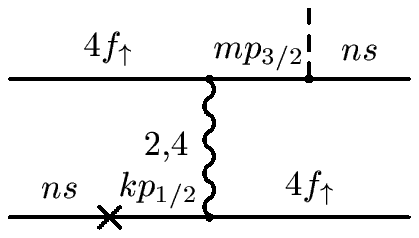,height=2cm}}\\
&4)& &5)& &6)\\

\multicolumn{1}{m{1cm}}{$-\quad 2 \times$} &
\multicolumn{1}{m{3.6cm}}{\epsfig{figure=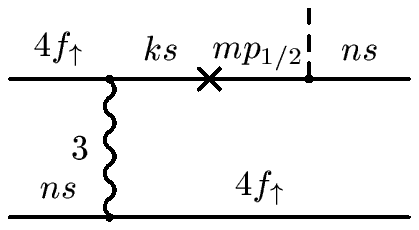,height=2cm}}&
\multicolumn{1}{m{1cm}}{$-\quad 2 \times$} &
\multicolumn{1}{m{3.6cm}}{\epsfig{figure=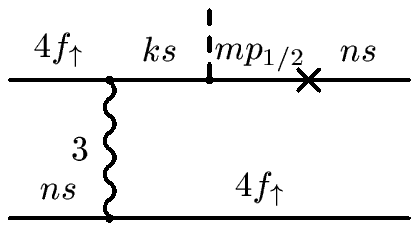,height=2cm}}&
&\\
&7)& &8)& &
\end{tabular}

\vspace{0.2cm}
\hspace{-3.0cm}
\begin{tabular}{l c c c c c c}

\multicolumn{1}{m{0.8cm}}{$+$} &
\multicolumn{1}{m{1.8cm}}{\epsfig{figure=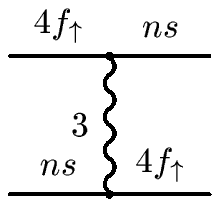,height=1.6cm}}&
\multicolumn{1}{m{0.8cm}}{$\times\quad\!\!\Bigg($} &
\multicolumn{1}{m{2.7cm}}{\epsfig{figure=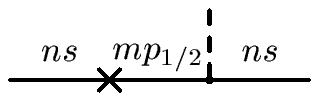,height=0.76cm}}&
\multicolumn{1}{m{0.7cm}}{$\quad\!\!\!+$} &
\multicolumn{1}{m{2.7cm}}{\epsfig{figure=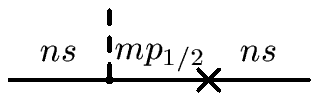,height=0.76cm}}&
\multicolumn{1}{m{1cm}}{$\Bigg)$}
\\& &9)& & & &
\end{tabular}

\vspace{0.1cm}

\begin{tabular}{c c c c c c}

\multicolumn{1}{m{1cm}}{$-\quad 2 \times$} &
\multicolumn{1}{m{3.6cm}}{\epsfig{figure=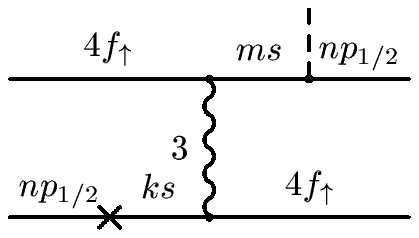,height=2cm}}&
\multicolumn{1}{m{1cm}}{$-\quad 2 \times$} &
\multicolumn{1}{m{3.6cm}}{\epsfig{figure=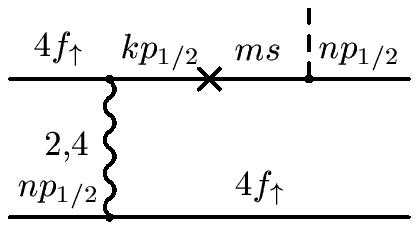,height=2cm}}&
\multicolumn{1}{m{1cm}}{$-\quad 2 \times$} &
\multicolumn{1}{m{3.6cm}}{\epsfig{figure=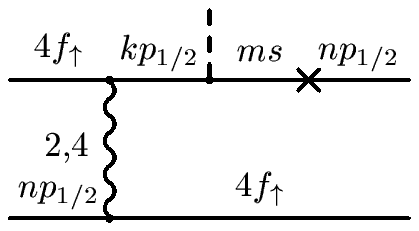,height=2cm}}\\
&10)& &11)& &12)\\

\multicolumn{1}{m{1cm}}{$-\quad 2 \times$} &
\multicolumn{1}{m{3.6cm}}{\epsfig{figure=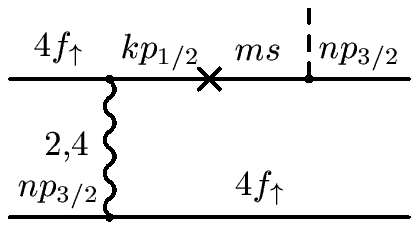,height=2cm}}&
\multicolumn{1}{m{1cm}}{$-\quad 2 \times$} &
\multicolumn{1}{m{3.6cm}}{\epsfig{figure=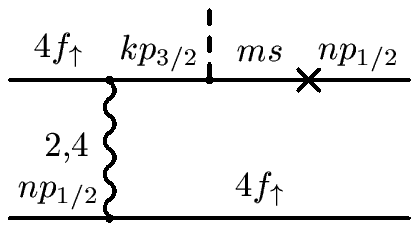,height=2cm}}&
&\\
&13)& &14)& &
\end{tabular}

\vspace{0.2cm}
\hspace{-3.0cm}
\begin{tabular}{l c c c c c c}

\multicolumn{1}{m{0.8cm}}{$+$} &
\multicolumn{1}{m{1.8cm}}{\epsfig{figure=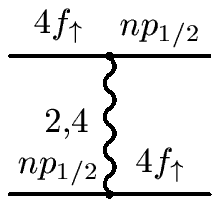,height=1.6cm}}&
\multicolumn{1}{m{0.8cm}}{$\times\quad\!\!\Bigg($} &
\multicolumn{1}{m{2.7cm}}{\epsfig{figure=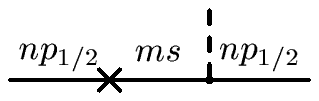,height=0.76cm}}&
\multicolumn{1}{m{0.7cm}}{$\quad\!\!\!+$} &
\multicolumn{1}{m{2.7cm}}{\epsfig{figure=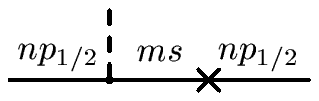,height=0.76cm}}&
\multicolumn{1}{m{1cm}}{$\Bigg)$}
\\
& &15)& & & &
\end{tabular}
\caption{Third order perturbation theory diagrams corresponding to
Eq.~(\ref{3d}). The cross denotes the T,P-odd interaction $V_d^r$ (\ref{pert1}),
the dashed line denotes the lattice deformation perturbation $V_x$ (\ref{pert2}),
and the wave line denotes the Coulomb interaction $V_C$ (\ref{pert3}). Multipolarity
of the Coulomb interaction is shown near the line.
Each diagram contributes with a coefficient shown
before the diagram.
Summation over {\it all} intermediate states $|k\rangle$ and $|m\rangle$ and
over {\it all filled} states $|n\rangle$ is assumed.}
\label{alld}
\end{figure}

Formula for the energy correction in the third order of perturbation theory 
reads, see e.g.~Ref.~\cite{LL}
\begin{equation}
\label{3d}
E_n^{(3)}=\sum_m {}^{'}\sum_k {}^{'}\frac{V_{nm}V_{mk}V_{kn}}{\hbar^2 
\omega_{mn} \omega_{kn}} - V_{nn}\sum_m {}^{'}\frac{|V_{nm}|^2}{\hbar^2 
\omega_{nm}^2},
\end{equation}
where
\begin{equation}
V=V_d^r+V_x+V_C \ .
\end{equation}
In Eq.~(\ref{3d}) we need to consider only the terms that are linear in each
of the operators $V_d^r$  (\ref{pert1}), $V_x$ (\ref{pert2}), and $V_C$ (\ref{pert3}).
There are 15 diagrams corresponding to Eq.~(\ref{3d}), these diagrams are
presented in Fig.~\ref{alld}. Each diagram contributes with a coefficient shown
before the diagram. Summation over {\it all} intermediate states $|k\rangle$ and 
$|m\rangle$ and over {\it all filled} states $|n\rangle$ is assumed.
The first four are the direct diagrams that correspond to the mechanism considered
in \cite{Simon}. All other diagrams are exchange ones and therefore they
contribute with sign (--). This is not so with respect to the disconnected 
exchange diagrams with brackets. These contributions correspond to the negative 
term in (\ref{3d}). In the exchange diagrams we account only for $s$-$p$ mixing by $V_d^r$,
contributions of higher angular momenta are neglected, see e.g.~Ref.~\cite{KL}. The 
deformation  (dashed line) is also attached only to $s$-$p$ lines because it is practically 
saturated by $6s$- and $6p$-states.
Since $V_d^r$ and $V_x$ are single particle operators we evaluate each diagram
solving equations for corresponding corrections. For example the first diagram
contains on the right top leg  the correction
\begin{equation}
\label{dp1}
\delta\psi_x = \sum_m \frac{\langle mp_{1/2}|V_x|ns\rangle}
{\epsilon_{ns}-\epsilon_{mp_{1/2}}}|mp_{1/2}\rangle \ .
\end{equation}
To evaluate the correction we do not use a direct summation, but instead
solve the equation
\begin{equation}
\label{dp2}
(H-\epsilon)\delta\psi_x = -V_x|ns\rangle \ , \ \ \ \epsilon =\epsilon_{ns}
\end{equation}
for each particular $|ns\rangle$ state. Here $H$ is the Dirac Hamiltonian
with potential (\ref{pot2}).
Similarly the bottom left leg of the diagram 5 is evaluated using
\begin{equation}
\label{dp3}
(H-\epsilon_{ns})\delta\psi_d = -V_d^r|ns\rangle
\end{equation}

\noindent
Diagrams with two single particle operators on the leg require a more careful treatment.
For example to evaluate the top right leg of the diagram 7 we first find $\delta\psi_x$
using Eq.~(\ref{dp2}). After that we calculate $\delta\psi_{dx}$ using
\begin{equation}
\label{dp4}
(H-\epsilon_{ns})\delta\psi_{dx} = -V_d^r\delta\psi_x+
\langle ns|V_d^r|\delta\psi_x\rangle |ns\rangle \ .
\end{equation}
The additional term $\langle ns|V_d^r|\delta\psi_x\rangle |ns\rangle$
in the right hand side of the equation is due
to the orthogonality condition $\langle \delta\psi_{dx}|ns\rangle=0$.
The corrections $\delta\psi_{dx}$ in diagrams 13 and 14 clearly do not require any 
additional terms in the corresponding equations.
Finally the disconnected diagrams (diagrams with brackets) which originate from
the second term in (\ref{3d}) contain the energy denominator squared. For example
the bracket in the diagram 9 is equal to $2\langle ns|V_d^r|\delta\phi_x\rangle$
where
\begin{equation}
\label{dp5}
\delta\phi_x = \sum_m \frac{\langle mp_{1/2}|V_x|ns\rangle}
{(\epsilon_{ns}-\epsilon_{mp_{1/2}})^2}|mp_{1/2}\rangle \ .
\end{equation}
To find $\delta\phi_x$ we calculate $\delta\psi_x$ at $\epsilon=\epsilon_{ns}\pm\delta$, 
see Eq.~(\ref{dp2}), and then find $\delta\phi_x$ using numerical differentiation
\begin{equation}
\delta\phi_x= -\left. \frac{\delta\psi_x(\epsilon_{ns}+\delta)-
\delta\psi_x(\epsilon_{ns}-\delta)}{2\delta}\right|_{\delta \to 0} \ .
\end{equation}

As a result of the calculations we get the following  T,P-odd energy correction
related to the displacement $x$
\begin{eqnarray}
\label{dx}
\Delta \epsilon (x) &=& -A \frac{x}{a_B} 
\left(\frac{d_e}{ea_B}\right)E_0 \ (\boldsymbol{n}_S\cdot \boldsymbol{n}_x)
\ ,\nonumber \\
A &=& -0.094 -0.159 +0.080 +0.133 +0.141 -0.295 -0.257\\
&-&\, 0.040 +0.610 -0.295 -0.009 -0.001 +0.440 -0.159\nonumber\\
&=&0.095\ .\nonumber
\end{eqnarray}
We remind that $a_B$ is the Bohr radius, and $e=|e|$ is the elementary charge,
so the ratio of the electron EDM $d_e$ to $e a_B$ is dimensionless.
The results in (\ref{dx}) is given is atomic units, $E_0=27.2eV$. 
The unit vectors are $\boldsymbol{n}_S=\boldsymbol{S}/S$, 
$\boldsymbol{n}_x=\boldsymbol{x}/x$, where $S=7/2$ is spin of Gd ion. Fourteen  terms in $A$
represent contributions of fifteen diagrams Fig.~\ref{alld}.
For diagrams 13 and 14 we present only the combined contribution ($+0.440$).
The point is that each of these two diagrams gives a very large contribution
inversely proportional to the fine structure splitting. However these large
contributions are canceled out in the sum of 13 and 14, so the sum remains
finite even at zero fine structure splitting. For this reason we do not calculate
13 and 14 separately.

\section{T,P-odd voltage across magnetically polarized sample of GDIG,
and magnetization of GGG in the external electric field}

In a magnetically polarized sample, according  to Eq.~(\ref{dx}),
each Gd$^{3+}$ ion can gain energy from a small distortion of the lattice.
The lattice has some stiffness and therefore the  variation of  energy
per Gd ion as a function of the displacement $x$ is of the form
\begin{equation}
\Delta \epsilon(x) = \frac{1}{2}K x^2 - A \frac{x}{a_B}
\frac{d_e}{ea_B}E_0,
\end{equation}
where $K$ is the effective elastic constant per Gd$^{3+}$ ion. Minimum
value of $\Delta \epsilon(x)$ corresponds to the shift
\begin{equation}
\label{xe}
\frac{x}{a_B} = A\frac{E_0}{K a_B^2}\frac{d_e}{ea_B},
\end{equation}
which is the new equilibrium position of Gd$^{3+}$ in the magnetically
polarized sample.

To find $x$ one needs to know $K$.
In Ref.~\cite{Simon} value of this constant has been estimated
using the known static dielectric constant of GdIG $\epsilon\approx 15$
and using a rather simplistic picture of the dielectric polarization.
In the present work we estimate $K$ using analysis of data on infrared
spectroscopy of garnets \cite{ir1,ir2}.
According to \cite{ir1,ir2} the so called ``N'' and the ``O'' infrared modes are
related to movement of Gd. The N mode in GdIG has energy
$\omega_N=213 cm^{-1}$. Energy of the O mode is unfortunately unknown.
However similar modes are known in  Yttrium iron garnet:
$\omega_N=208.8 cm^{-1}$, $\omega_O=144 cm^{-1}$. Assuming the same ratio
$\omega_N/\omega_O$ we find for GdIG: $\omega_O=147 cm^{-1}$.
Movement of Gd is not a normal mode of the lattice, however the splitting
between $\omega_N$ and $\omega_O$ is not large. Therefore to find
the effective frequency we average between the modes
\begin{equation}
\label{ono}
\omega=\frac{1}{2}(\omega_N+\omega_O)=180 cm^{-1}.
\end{equation}
The elastic constant is equal to
\begin{equation}
\label{km}
K = \mu \omega^2,
\end{equation}
where $\mu$ is the reduced mass corresponding to movement of Gd with
respect to the lattice  of the garnet Gd$_3$Fe$_5$O$_{12}$,
\begin{equation}
\frac{1}{\mu} = \frac{1}{M_{Gd}}+\frac{3}{5M_{Fe}+12M_O}, \ \ \to \ \ \mu=77.
\end{equation}
Together with (\ref{ono}) and (\ref{km}) this gives
\begin{equation}
\label{km1}
K=0.095\frac{E_0}{a_B^2}.
\end{equation}
Accidentally the dimensionless coefficient in this equation is the same as
that in Eq.~(\ref{dx}). The value (\ref{km1}) is by factor 2.35 smaller than 
that obtained in Ref.~\cite{Simon} from the dielectric constant.
We believe that the present analysis is more reliable and gives the more
accurate value of the elastic constant.
Further analysis of the lattice vibrations and a new
data on the infrared absorption would be very helpful for precise
determination of the elastic constant.

Using Eqs.~(\ref{xe}) and (\ref{km1}) we obtain the  value
of the displacement induced by the electron EDM, $x \approx d_e/e$.
Macroscopic electric polarization arising from the shift of Gd ions in the
sample is given by $P = 3exn_{Gd}$, where $n_{Gd}=1.235 \times 10^{22} \ cm^{-3}$
is number density of Gd in GdIG. Hence, the electric field in the sample is
\begin{equation}
E = -4\pi P \approx 12\pi n_{Gd}  d_e = - 1.1 \times 10^{-10}V/cm.
\end{equation}
The numerical value corresponds to the current upper limit on the
electron EDM (\ref{Tl}). For a $10cm$ sample this gives a voltage
$\Delta V = 1.1\times 10^{-9} V$. This value is a factor 3 greater
than that obtained in \cite{Simon}. The difference is mainly due to the
different elastic constant.

Another effect is magnetization in the external electric field.
A candidate for such experiment is gadolinium gallium garnet (GGG),
see Ref.~\cite{Lam}. There is no need to repeat the calculation
\cite{Simon}, we need only to rescale the value.  According to the 
present work the value of
$A$ in Eq.~(\ref{dx}) is a factor 1.27 larger than that from \cite{Simon}
and the elastic constant is a factor 2.35 smaller than that
from \cite{Simon} (we assume that the elastic constants for GdIG and GGG
coincide). Therefore, the energy shift of Gd$^{3+}$ ion in GGG in the
external electric field is 
\begin{equation}
\label{de}
\Delta \epsilon = (\boldsymbol{n}_E\cdot \boldsymbol{n}_S) \ 5.7 \times 10^{-22}eV \ ,
\end{equation}
where $\boldsymbol{n}_E$ is the unit vector along the electric field, and
$\boldsymbol{n}_S$ is the unit vector along the ion spin. The value corresponds
to the current upper limit on $d_e$ (\ref{Tl}) and the electric field
inside the sample $E=10 \ kV/cm$.
The energy shift (\ref{de}) leads to the macroscopic magnetization of the
sample. The magnetization depends also on temperature and internal magnetic
interactions in the compound. We do not discuss these points here.
According to estimates \cite{Lam} the magnetization due to the energy
shift $\sim 10^{-22}eV$ can be measured and moreover the prospects for 
improvement of sensitivity are very good.

\section{Conclusions}
In the present work we have calculated  the T,P-odd effects 
induced by the electron electric dipole moment  (EDM) in  gadolinium gallium
garnet, Gd$_3$Ga$_5$O$_{12}$ (GGG), and gadolinium iron garnet 
Gd$_3$Fe$_5$O$_{12}$ (GdIG).   Both GdIG and GGG have uncompensated electron spins on 
Gd$^{3+}$ ions. There are two possibilities to probe the electron EDM.
The first one is to polarize magnetically the electron spins and to measure
the induced voltage across the sample. According to the present calculation
at the current limitation on the electron EDM, (\ref{Tl}), the induced voltage 
across a $10cm$ sample is $1.1\times 10^{-9}V$.
Another possibility is to apply an electric field to the unpolarized sample.
This leads to the spin-dependent energy shift of each Gd ion $\Delta \epsilon =
5.7 \times 10^{-22} eV$ at $E=10kV/cm$. This can be 
measured via macroscopic magnetization of the sample.

In the present work we generally followed the path of \cite{Simon}. However the 
electronic part of the problem has been considered much more accurately with account of
exchange diagrams which have been neglected in  \cite{Simon}. Surprisingly, because of
accidental compensations, the final electronic effective Hamiltonian proved to be
close to that from \cite{Simon}. Nevertheless the observable effects are by 3 times
bigger than that in \cite{Simon}. The main reason for this is in the different lattice
elastic constant.  To determine value of the constant in the present work we have used
data on infrared absorption of garnets. A new data on the infrared absorption and
further analysis of the lattice vibrations would be very helpful for more accurate
calculations of observable effects. However the main problem which bothers us is the
electronic part.  In the present work we have used a ``jelly model'' smearing  eight
oxygen ions surrounding  gadolinium ion over a spherical shell. 
It is a sensible simplification. However we found strong compensations between 
contributions of different exchange diagrams. Clearly the compensations are related to 
the Schiff theorem. In this situation any simplification causes questions and, in
our opinion, a further calculation of the effect with account of real 3D geometry is 
necessary.


\end{document}